\documentclass[12pt,a4paper]{article}
\usepackage{wasysym}
\usepackage{a4wide}
\usepackage{epsfig}
\usepackage{subfigure}
\usepackage{revsymb}
\usepackage{latexsym}
\usepackage{amssymb}

\begin{document}

\begin{center}
{\large\bf Final State Interactions in $B\to\pi\pi,\,\pi K$ decays

 and constrains on $\gamma$}

\bigskip

\bigskip

Erika Alvarez$^{1,2}$, Jean Pestieau$^2$

\bigskip

$^1${\it Instituto de F\' isica, Universidad Nacional Aut\'onoma de M\'exico,

 Apartado Postal 20364, 01000, M\'exico D.F., M\'exico

\bigskip

$^2$ Institut de Physique Th\'eorique, Universit\'e catholique de Louvain 

 B-1348 Louvain-la-Neuve, Belgium}

\bigskip

\bigskip

{\bf Abstract}

\end{center}

We apply the quasi-elastic assumption of the strong final state interactions to describe the $B\to\pi\pi$ and $B\to\pi K$ branching ratios and CP asymmetries and derive bounds on $\gamma$. We find that small electroweak penguins contributions and values of $\gamma$ larger than in standard fit results are favored by data.

\pagebreak

\section{Introduction}
The strong phases of the $B\to\pi\pi$ and $B\to\pi K$ decay amplitudes have been usually treated assuming that the strong interactions are dominated by short distance perturbative effects. According to this the 
strong phases can be obtained perturbatively from penguin and tree 
diagrams and are then generalized to describe hadronic strong 
phases to all orders. However, in \cite{gerard} it was shown the lack of compatibility between the strong phases calculated perturbatively, and those of the isospin basis, where phases are contemplated at all scales (soft and hard).

The isospin symmetry is a very good approximation of the hadronic world. For example, the $K\to\pi\pi$ decay amplitudes are treated on the basis of this symmetry of the strong interactions. For $B$ decays we still have isospin invariance but now there are many modes apart from the elastic ones that can be reached with rescattering of final states (FSI). However, for the case of $B$ mesons inelasticity is not expected to be significant as far as it becomes important for large angular momentum waves. Therefore we will work in the quasi-elastic limit, where inelastic channels are neglected in rescattering.

 The Regge model predicts that the quasi-elastic strong phases are small, and so the CP asymmetries must be necessarily small. Experimentally, the $\pi^-K^+$ asymmetry turns out to be large, in contradiction with the hypothesis. Nevertheless, this mode involves a charged pion and 
corrections due to radiation of photons can play an important role in 
quoting precise results from experiments. 

In this paper we use the quasi-elastic predictions for FSI in $B\to K\pi,\ \pi\pi$ to derive bounds on the angle $\gamma$ of the unitary triangle. We give an overview of the Watson Theorem and on the concepts of bare and rescattered 
amplitudes, then apply these principles to parameterize the decay 
amplitudes in terms of the quark and isospin amplitudes. Using the 
experimental branching ratios, which are more precisely measured than the 
CP asymmetries, we give an estimate for the range of $\gamma$ and of the electroweak penguins contribution, then we give predictions for the asymmetries. We 
finally present our conclusions.

\section{Hadronic strong phases}

 The Watson Theorem \cite{wolfenstein2}, \cite{christopher} tell us that the decay amplitudes factorize into the product of the direct or ``bare'' weak amplitudes, $A^b$, and the rescattering factors, $\sqrt S$,
$$A_i(B\to P_i)=\sqrt S_{ij} A^b_j(B\to P_j),\quad \overline A_i(\overline B\to \overline P_i)=\sqrt S_{ij} A^{b*}_j(B\to P_j),$$
where as usual the $CP$ conjugated amplitude is given the same strong 
phase $S_{ij}$ which denotes here the  matrix element of the 
 $P_i\to P_j$ strong interaction. If we factorize the weak phase in the amplitude $A^b$, we do not have other strong phases induced by perturbative calculations, and the complete hadronic phases to 
all orders are included in the  $\sqrt 
S$ factors. 

If we were in the elastic case, $\sqrt S$ would be diagonal in the isospin basis:
$$\sqrt S_{el}\!=\!\left(\begin{array}{cc}e^{i\delta_{3/2}} & \\ & e^{i\delta_{1/2}}\end{array}\right).$$
Thus, for the case of $B\to\pi K$ decay amplitudes, we would have 
 $$A_{I}\to A_{I}e^{i\delta_{I}},\quad \overline A_{I}\to A^*_{I}e^{i\delta_{I}},\quad I=3/2,1/2,$$
or in the physical basis
$$\left(\!\begin{array}{c}A_{\pi^+K^0}\\A_{\pi^0 K^+}\end{array}\!\right)\!=\!\left(\!\begin{array}{cc} -\frac{1}{\sqrt 3} & \sqrt{\frac{2}{3}}\\ \sqrt{\frac{2}{3}} & \frac{1}{\sqrt 3}\end{array}\!\right)\!\left(\!\begin{array}{c}A_{3/2}e^{i\delta_{3/2}}\\A_{1/2}e^{i\delta_{1/2}}\end{array}\!\right),\quad \left(\!\begin{array}{c}\overline A_{\pi^-\overline K^0}\\\overline A_{\pi^0 K^-}\end{array}\!\right)\!=\!\left(\!\begin{array}{cc} -\frac{1}{\sqrt 3} & \sqrt{\frac{2}{3}}\\ \sqrt{\frac{2}{3}} & \frac{1}{\sqrt 3}\end{array}\!\right)\!\left(\!\begin{array}{c}A^*_{3/2}e^{i\delta_{3/2}}\\A^*_{1/2}e^{i\delta_{1/2}}\end{array}\!\right),$$
with analogous expressions for the other channels $A_{\pi^0K^0}$ and $A_{\pi^-K^+}$ that are reached by $B^0$ decays.

At the $B$ mass, however, we have the more general inelastic case, which 
is described in the isospin symmetric case as
$$\left(\!\begin{array}{c}A_{\pi^+K^0}\\A_{\pi^0 K^+}\end{array}\!\right)\!=\!\left(\!\begin{array}{cc} -\frac{1}{\sqrt 3} & \sqrt{\frac{2}{3}}\\ \sqrt{\frac{2}{3}} & \frac{1}{\sqrt 3}\end{array}\!\right)\!\left(\!\begin{array}{c}A_{3/2}\sqrt{\eta}_{3/2}e^{i\delta_{3/2}}+\sum_{j\neq \pi K}\sqrt S^{I=3/2}_{\pi K,j}A_{3/2}^j\\A_{1/2}\sqrt{\eta}_{1/2}e^{i\delta_{1/2}}+\sum_{j\neq \pi K}\sqrt S^{I=1/2}_{\pi K,j}A_{1/2}^j\end{array}\!\right)$$
where we have replaced $\sqrt S_{ii}\!=\!e^{i\delta_I}\!\to\!\sqrt{\eta}_Ie^{i\delta_I}$ for the 
elastic FSI and have also incorporated the inelastic channels ($j\not = 
\pi K$). The elastic parameters $\eta$ and 
$\delta$, can be calculated using the Regge model of strong 
interactions.

In the Regge model, the high energy amplitudes in  the $s$-channel can be parametrized in terms of trajectories exchanges in the small $t$ region.  A 
Regge Model to describe $B\to \pi K,\ \pi\pi$ decays was 
proposed in \cite{jeanpion} and \cite{jeankaon}, but this assumed the complete 
elasticity of FSI as it was later remarked in \cite{wolfenstein3}. Taking 
into account this remark and using the 
relations between the isospin amplitudes in the $s$ and $t$ channels 
$$\left(\begin{array}{c} A_0^s \\  A_1^s \\  A_2^s \end{array}\right)=\left(\begin{array}{ccc} 1/3 & 1 & 5/3 \\  1/3 & 1/2 & -5/6 \\  1/3 & -1/2 & 1/6 \end{array}\right)\left(\begin{array}{c} A_0^t \\  A_1^t \\  A_2^t \end{array}\right),\quad \left(\begin{array}{c} A_{1/2}^s \\  A_{3/2}^s \end{array}\right)=\left(\begin{array}{cc} 1/\sqrt{6} & 1  \\  1/\sqrt{6} & -1/2  \end{array}\right)\left(\begin{array}{c} A_0^t \\  A_1^t \end{array}\right)$$
we now find:
 $$\eta_{I=1/2}^{\pi K}(m_B)=0.72,\ \eta_{I=3/2}^{\pi K}(m_B)=0.71,$$ 
 $$\eta_{I=0}^{\pi\pi}(m_B)=\eta_{I=2}^{\pi\pi}(m_B)=0.64.$$ 

For these results we made a fit of nucleon-nucleon and pion-nucleon total 
cross sections and made use of factorization. $f_2$ and $\rho$ trajectories were considered in t channel in addition to the Pomeron exchange, for both, $\pi\!-\!\pi$ and $\pi\!-\!K$ modes. $K^*$ and $K^{**}$ trajectories were neglected in the small $-\!u$ region in the case of $\pi\!-\!K$, because of their small size.
The equality of the $\eta$ factors in each mode follows from the pomeron 
dominance over the rest of trajectories contributing to the $A_0^t$ 
amplitude at the $B$ mass scale. 
 As a consequence the strong phases are almost the same (and close to 
zero because the pomeron is almost imaginary), so that the differences of 
the strong phases would be very tiny and it becomes difficult to give its 
value with accuracy.

Since the factors $\sqrt{\eta}$ are close to unity, we 
can take the {\it quasi-elastic assumption} \cite{german}, that consists 
in the  cancellation of the inelastic channels, which means that
$\sum_{j\neq \pi K}\sqrt S^{I}_{\pi K,j}A_{I}^j\to 0$.

\section{Amplitudes decomposition}

Let us consider the $B\to\pi K$ decay amplitudes. Once we absorb the elastic 
parameters $\sqrt{\eta}_I$ into the definitions of physical amplitudes, they can be 
written in terms of the isospin amplitudes as follows
$$A_{\pi^+K^0}\!=\!-\frac{A_{3/2}}{\sqrt{3}}e^{i\delta_{3/2}}\!+\!\sqrt{\frac{2}{3}}(A^{\prime}_{1/2}\!+\!A_{1/2})e^{i\delta_{1/2}},\quad A_{\pi^0K^+}\!=\!\sqrt{\frac{2}{3}}A_{3/2}e^{i\delta_{3/2}}\!+\!\frac{1}{\sqrt 3}(A^{\prime}_{1/2}\!+\!A_{1/2})e^{i\delta_{1/2}},$$
$$A_{\pi^0K^0}\!=\!\sqrt{\frac{2}{3}}A_{3/2}e^{i\delta_{3/2}}\!-\!\frac{1}{\sqrt 3}(A^{\prime}_{1/2}\!-\!A_{1/2})e^{i\delta_{1/2}},\quad A_{\pi^-K^+}\!=\!\frac{A_{3/2}}{\sqrt{3}}e^{i\delta_{3/2}}\!+\!\sqrt{\frac{2}{3}}(A^{\prime}_{1/2}\!-\!A_{1/2})e^{i\delta_{1/2}},$$
where we have defined the isospin amplitudes
$$A_{3/2}=-\sqrt{\frac{2}{3}}\langle 3/2|H_{\Delta I=1}|1/2\rangle,\quad A_{1/2}=-\frac{1}{\sqrt 3}\langle 1/2| H_{\Delta I=1}|1/2\rangle,\quad A^{\prime}_{1/2}=\langle 1/2 |H_{\Delta I=0}|1/2\rangle .$$

On the other hand, we can express the bare or direct weak amplitudes in terms of the leading  $T$ ({\em tree}), $P$ ({\em QCD-penguin}), $C$ ({\em color suppressed}) and $P^{EW}$ ({\em EW-penguin}) quark diagrams as follows  
 $$A^b_{\pi^+K^0}=P\lambda^s_t,\qquad A^b_{\pi^0K^+}=\frac{1}{\sqrt 2}\left((T+C)\lambda^s_u+(P+P^{EW})\lambda^s_t\right),$$
$$ A^b_{\pi^0K^0}=\frac{1}{\sqrt 2}\left(C\lambda^s_u-(P-P^{EW})\lambda^s_t\right),\qquad A^b_{\pi^-K^+}=\left(T\lambda^s_u+P\lambda^s_t\right),$$
where we have defined $\lambda_t^s=V_{ts}V_{tb}^*\approx -A\lambda^2$ and 
$\lambda_u^s=V_{us}V_{ub}^*\approx AR_u\lambda^4e^{i\gamma}$, with 
$R_u\!=\!|\rho\!-\!i\eta|$ in the Wolfenstein parameterization.

When the strong phases are turned off in the isospin amplitudes, we can identify the 
following relations between the quark and the isospin amplitudes, 
$$A_{3/2}=\frac{1}{\sqrt 3}\left[(T+C)\lambda_u^s+P^{EW}\lambda_t^s\right],\quad A_{1/2}=\frac{1}{2\sqrt 6}\left[(-T+2C)\lambda_u^s+2P^{EW}\lambda_t^s\right],$$$$A'_{1/2}=\sqrt{\frac{3}{2}}\left(\frac{T}{2}\lambda^s_u+P\lambda^s_t\right).$$
Thus, using such identities we find the physical amplitudes (including FSI) in 
terms of quark diagrams 
$$A_{\pi^+K^0}=P\lambda^s_t+\frac{1}{3}\left((T+C)\lambda^s_u+P^{EW}\lambda^s_t\right)(1-e^{i\delta_{\pi K}}),$$
$$A_{\pi^0K^+}=\frac{1}{\sqrt 2}\left((T+C)\lambda^s_u+(P+P^{EW})\lambda^s_t\right)-\frac{\sqrt 2}{3}\left((T+C)\lambda^s_u+P^{EW}\lambda^s_t\right)(1-e^{i\delta_{\pi K}}),$$
$$A_{\pi^0K^0}=\frac{1}{\sqrt 2}\left(C\lambda^s_u-(P-P^{EW})\lambda^s_t\right)-\frac{\sqrt 2}{3}\left((T+C)\lambda^s_u+P^{EW}\lambda^s_t\right)(1-e^{i\delta_{\pi K}}),$$
$$A_{\pi^-K^+}=\left(T\lambda^s_u+P\lambda^s_t\right)-\frac{1}{3}\left((T+C)\lambda^s_u+P^{EW}\lambda^s_t\right)(1-e^{i\delta_{\pi K}}),$$
where the strong phase $\delta_{1/2}$ was absorbed into the definitions of the 
amplitudes and $\delta_{\pi K}\! \equiv \!\delta_{3/2}\!-\!\delta_{1/2}$.

In the case of $B\to\pi\pi$ decays, we follow the same steps. Here, the physical amplitudes in terms of the isospin amplitudes are given by
$$A_{\pi^+\pi^0}=\sqrt{\frac{3}{2}}A_2 e^{i\delta_2},\quad A_{\pi^+\pi^-}=\sqrt{\frac{2}{3}}A_0e^{i\delta_0}+\frac{A_2}{\sqrt 3}e^{i\delta_2},\quad A_{\pi^0\pi^0}=\frac{A_0}{\sqrt 3}e^{i\delta_0}-\sqrt{\frac{2}{3}}A_2 e^{i\delta_2},$$
where
$$A_{2}=-\frac{1}{\sqrt{2}}\langle 2|H_{\Delta I=3/2}|1/2\rangle,\quad A_{0}=-\frac{1}{\sqrt 2}\langle 0| H_{\Delta I=1/2}|1/2\rangle .$$

In terms of the leading quark diagrams, the bare weak amplitudes are
$$A_{\pi^+\pi^0}^b=\frac{1}{\sqrt 2}(T+C)\lambda_u^d,\quad A_{\pi^+\pi^-}^b=T\lambda_u^d+P\lambda_t^d,\quad A_{\pi^0\pi^0}^b=\frac{1}{\sqrt 2}\left(-C\lambda_u^d+P\lambda_t^d\right),$$
where  $\lambda_u^d=V^*_{ub}V_{ud}\approx AR_u\lambda^3e^{i\gamma}$ and $\lambda_t^d=V^*_{tb}V_{td}\approx AR_t\lambda^3e^{-i\beta}$, with $R_t\!=\!|1\!-\!\rho\!-\!i\eta|$.

In the absence of the strong phases, we have the following relations between the isospin and quark amplitudes:
$$A_2=\frac{1}{\sqrt 3}(T+C)\lambda_u, \qquad A_0=\left(\sqrt{\frac{2}{3}}T-\frac{C}{\sqrt 6}\right)\lambda_u+\sqrt{\frac{3}{2}}P\lambda_t$$
Using these relations, the total rescattered amplitudes in terms of quarks 
diagrams are 
$$A_{\pi^+\pi^0}=\frac{1}{\sqrt 2}(T+C)\lambda_u^d e^{i\delta_{\pi\pi}},$$
$$ A_{\pi^+\pi^-}=(T\lambda_u^d+P\lambda_t^d)+\frac{1}{3}(T+C)\lambda_u^d\left ( e^{i\delta_{\pi\pi}}-1\right ),$$
$$A_{\pi^0\pi^0}=\frac{1}{\sqrt 2}\left(-C\lambda_u^d+P\lambda_t^d\right)-\frac{\sqrt 2}{3}(T+C)\lambda_u^d\left ( e^{i\delta_{\pi\pi}}-1\right ),$$
where $\delta_{\pi\pi}\!=\!\delta_2-\delta_0$.

Measurements of the $B\!\to\!\pi K$ and $B\!\to\!\pi \pi$ branching ratios have been 
reported in refs. \cite{CLEObranching}-\cite{Bellepiceroasim} (expressed in units of  
$10^{-6}$ in Table 1).

 $$\begin{array}{|ccccc|}
\multicolumn{5}{l}{\mbox{TABLE 1: Branching ratios of }B\to\pi K,\,\pi\pi\mbox{ decays}}\\\hline\hline
 & \mbox{BaBar}&\mbox{Belle} &\mbox{CLEO} & \mbox{average} \\\hline
\overline{{\cal B}}(B^+\!\to\!\pi^+ K^0) & 26.0\!\pm\! 1.6 & 22.0\!\pm\! 2.2 & 18.8^{\!+\!4.3}_{\!-\!3.8} & 24.1\!\pm\! 1.3 \\ 
\overline{{\cal B}}(B^+\!\to\!\pi^0K^+) &12.0\!\pm\! 0.9 & 12.0^{\!+\!1.8}_{\!-\!1.6} & 12.9^{\!+\!2.7}_{\!-\!2.5} & 12.1\!\pm\! 0.8  \\ 
\overline{{\cal B}}(B^0\!\to\!\pi^0 K^0) & 11.4\!\pm\! 1.1 & 11.7\!\pm\! 2.6 & 12.8^{\!+\!4.3}_{\!-\!3.6} & 11.5\!\pm\! 1.0 \\
\overline{{\cal B}}(B^0\!\to\!\pi^- K^+) & 19.2\!\pm\! 0.85 & 18.5\!\pm\! 1.2 & 18.0^{\!+\!2.6}_{\!-\!2.8} & 18.9\!\pm\! 0.7  \\\hline 
\overline{{\cal B}}(B^+\!\to\!\pi^+\pi^0) & 5.8\!\pm\! 0.7 & 5.0\!\pm\! 1.3 & 4.6^{\!+\!1.9}_{\!-\!1.7} & 5.5\!\pm\! 0.6 \\ 
\overline{{\cal B}}(B^0\!\to\!\pi^+\pi^-) & 5.5\!\pm\! 0.5 & 4.4\!\pm\! 0.67 & 4.5^{\!+\!1.5}_{\!-\!1.3} & 5.0\!\pm\! 0.4 \\ 
\overline{{\cal B}}(B^0\!\to\!\pi^0\pi^0) & 1.17\!\pm\! 0.335 & 2.32^{\!+\!0.45}_{\!-\!0.58} & <4.4 &1.45\!\pm\! 0.29\\
\hline \end{array}$$

Corresponding measurements of the CP asymmetries are much less precise 
\cite{BabarKmas}-\cite{Bellepimasasim} and are summarized in Table 2.
 $$\begin{array}{|ccccc|}
\multicolumn{5}{l}{\mbox{TABLE 2: CP Asymmetries of }B\to\pi K,\,\pi\pi\mbox{ decays}}\\\hline\hline
 &\mbox{BaBar}&\mbox{Belle}&\mbox{CLEO}&\mbox{average}  \\
\hline
{\cal A}_{\pi^-\!K^+} & - 0.133\!\pm\! 0.031 & - 0.113\!\pm\! 0.022 & - 0.04\!\pm\! 0.16 & - 0.115\pm 0.018 \\
{\cal A}_{\pi^+\!K^0} & - 0.087\!\pm\! 0.047 & 0.05\!\pm\! 0.05 & 0.18\!\pm\! 0.24 &- 0.02\!\pm\! 0.034 \\
{\cal A}_{\pi^0\!K^+} & 0.06\!\pm\! 0.06 & 0.04\!\pm\! 0.05 &- 0.29\!\pm\! 0.23&0.04\!\pm\! 0.04 \\
{\cal A}_{\pi^0\!K_S} & - 0.06\!\pm\! 0.19 & - 0.12\!\pm\! 0.21 &- & - 0.09\!\pm\! 0.14 
 \\\hline
 C_{\pi^+\!\pi^-} & - 0.09\!\pm\! 0.15 & - 0.56\!\pm\! 0.13  & - & - 0.37\!\pm\! 0.10\\
C_{\pi^0\!\pi^0}&- 0.12\!\pm\! 0.56& - 0.44^{+\!0.56}_{-\!0.55}&-  &- 0.28^{\!+\!0.40}_{\!-\!0.39}\\\hline
S_{\pi^+\!\pi^-} &  - 0.30\!\pm\! 0.17 &  - 0.67\!\pm\! 0.17 & - &- 0.50\!\pm\! 0.12 \\
\hline
\end{array}$$

 There are some noticeable differences between BaBar and Belle results, namely the differences of the branching ratios $\overline{{\cal B}}(B^+\!\to\!\pi^+K^0)$ and $\overline{{\cal B}}(B^0\!\to\!\pi^0\pi^0)$ are $1.5\sigma$ and $2\sigma$ away, respectively, while the differences of the asymmetries $S_{\pi^+\pi^-}$, ${\cal A}_{\pi^+K^0}$ and $C_{\pi^+\pi^-}$ differs by $1.5\sigma$, $2\sigma$ and $2.3\sigma$. Therefore, we prefer to do a separated analysis for both sets of data.

\section{Constraints from the branching ratios}

We define the following $6$ independent ratios of the CP-averaged $B\!\to\!\pi\pi,\pi K$ branching ratios 
$$R_c=\frac{2\overline{{\cal B}}(B^+\!\to\!\pi^0K^+)}{\overline{{\cal B}}(B^+\!\to\!\pi^+ K^0)},\quad
R_n=\frac{\overline{{\cal B}}(B^0\!\to\!\pi^-K^+)}{2\overline{{\cal B}}(B^0\!\to\!\pi^0K^0)},$$
$$ R_m=\frac{2\tau_+}{\tau_0}\frac{\overline{{\cal B}}(B^0\!\to\! \pi^0 K^0)}{\overline{{\cal B}}(B^+\!\to\!\pi^+K^0)},\quad
R_{\pi K}=\frac{2\overline{{\cal B}}(B^+\!\to\!\pi^+\pi^0)}{\overline{{\cal B}}(B^+\!\to\!\pi^+ K^0)},$$
$$ R_{\pi\pi}=\frac{\tau_+}{\tau_0}\frac{(\overline{{\cal B}}(B^0\!\to\!\pi^+\pi^-)+\overline{{\cal B}}(B^0\!\to\!\pi^0\pi^0))}{\overline{{\cal B}}(B^+\to\pi^+\pi^0)},\quad
 R_{0}=\frac{\tau_+}{\tau_0}\frac{\overline{{\cal B}}(B^0\!\to\!\pi^+\pi^-)}{\overline{{\cal B}}(B^+\!\to\!\pi^+\pi^0)},$$
where $\tau_+/\tau_0=1.076\pm 0.008$ \cite{HFAG} is the ratio of the $B^+$ and $B^0$ lifetimes. $R_c$ and $R_n$ were already introduced in \cite{buras}.
 
From Table 1 we obtain the following table
$$\begin{array}{|c|cccccc|}
\multicolumn{1}{c}{\mbox{exp}} & \multicolumn{1}{c}{R_{\pi K}} & \multicolumn{1}{c}{R_{c}} & \multicolumn{1}{c}{R_{m}} & \multicolumn{1}{c}{R_n} & \multicolumn{1}{c}{R_{\pi\pi}} &\multicolumn{1}{c}{R_{0}}\\\hline
 \mbox{BaBar} & 0.45\pm 0.09 & 0.92\pm 0.09 & 0.94\pm 0.11 & 0.84\pm 0.09 & 1.24\pm 0.19 & 1.02\pm 0.15\\\mbox{Belle} & 0.45\pm 0.19 & 1.09\pm 0.19 & 1.14\pm 0.28 & 0.79\pm 0.18 &1.45\pm 0.42& 0.95\pm 0.29\\\mbox{{\small average}} &0.46\pm 0.06 & 1.00\pm 0.08 & 1.03\pm 0.105 & 0.82\pm 0.08 & 1.26\pm 0.17 &0.98\pm 0.13 \\\hline\end{array}$$
where CLEO data has been included in the average.

From $R_{\pi K}$ we can obtain the tree (color allowed and suppressed) to penguin ratio. Defining
$$r=\frac{T|\lambda^s_{u}|}{P|\lambda^s_{t}|}=\frac{T\lambda^2 R_u}{P},\qquad r_c=\frac{(T+C)\lambda^2 R_u}{P},$$
$R_{\pi K}$ is written as
\begin{equation}\label{TCP}
R_{\pi K}=\frac{f_{\pi}^2}{f_K^2}\frac{(T+C)^2|\lambda_u^d|^2}{P^2|\lambda_t^s|^2}=\frac{f_{\pi}^2}{f_K^2}\frac{(T+C)^2(\lambda R_u)^2}{P^2}=\frac{f_{\pi}^2}{f_K^2}\frac{r_c^2}{\lambda^2}\quad\Rightarrow\quad r_c=\lambda \frac{f_K}{f_{\pi}}\sqrt{R_{\pi K}},
\end{equation}
where the ratio $f_{K}/f_{\pi}\!=\!1.2$ gives an estimate of the size of the $SU(3)$ symmetry breaking. Using the experimental value of $R_{\pi K}$ and $\lambda\!=\!\sin\theta_C\!=\!0.226$ we find $r_c\!=\!0.18\!\pm\! 0.01$. This result is obtained by neglecting the FSI corrections to $B^+\to\pi^+K^0$. 

The next step is to find the ratio $C/T$ by adding the information of the $B\!\to\!\pi\pi$ modes. In the quasi-elastic case
the final state interactions in  $B\to\pi^+\pi^-$ and $B\to\pi^0\pi^0$ cancel each other in their ratio and we have 

$$R_{\pi\pi}=\frac{2+3\left(\frac{P R_t}{T R_u}\right)^2+(C/T)^2+2\frac{P R_t}{T R_u}\left(\frac{C}{T}-2\right)\cos\alpha}{(1+C/T)^2}.$$
Since from Equation (\ref{TCP})
\begin{equation}\label{PT}
\frac{P}{T}=\frac{(1+C/T)\lambda R_u}{f_K/f_{\pi}\sqrt{R_{\pi K}}}
\end{equation}
 using the sine law ($R_t=\sin\gamma/\sin\alpha$ and $R_u=\sin\beta/\sin\alpha$) we can express the ratio $C/T$ in terms of $\gamma$ and $\beta$ as follows
\begin{equation}\label{CT}
\frac{C}{T}=\frac{-(A+B/2)+\sqrt{(3/2 B)^2+3 A-2}}{A-B-1}
\end{equation}
where
$$A=R_{\pi\pi}-\frac{3\lambda^2 \sin^2\gamma}{(f_K/f_{\pi})^2R_{\pi K}\sin^2(\pi-\gamma-\beta)},\quad B=\frac{2\lambda\sin\gamma}{(f_K/f_{\pi})\sqrt{R_{\pi K}}\tan(\pi-\gamma-\beta)}.$$

\begin{figure}
\centering{
\mbox{\subfigure{\epsfig{file=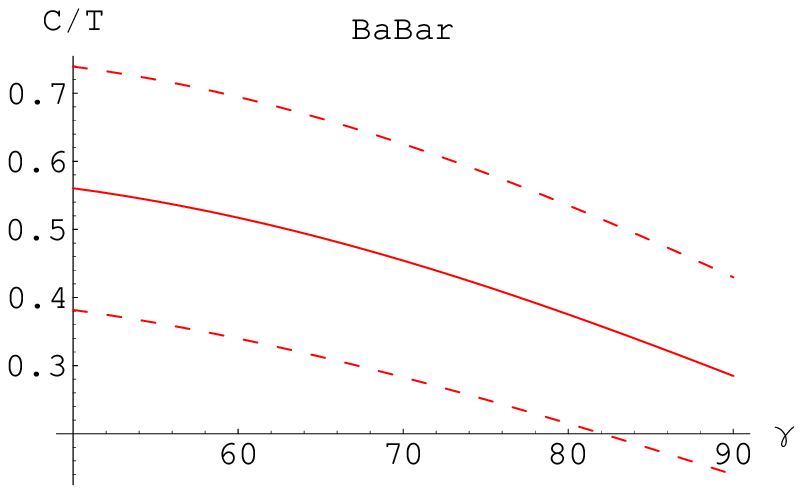,width=6cm,height=4cm}}\quad\subfigure{\epsfig{file=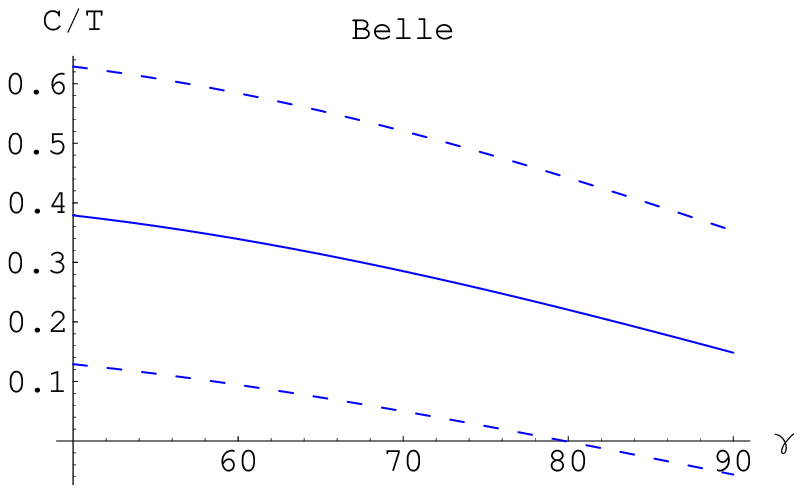,width=6cm,height=4cm}}}
}
\caption{ $C/T$ vs $\gamma$ extracted from ratios of B decay widths.}
\label{padre}
\end{figure}

In Figure (\ref{padre}) we plot $C/T$ as a function of $\gamma$, using the experimental value $\beta$, $\beta\!=\!(21.70\pm 1.26)^{\circ}$ \cite{ckm}. We 
observe that in order to have the naive expectation $C/T\sim 
0.3$ from  BaBar data we require $\gamma\gtrsim 68^{\circ}$, namely a value that is above the upper limit of the standard fit, which taken at its $1\sigma$ interval reads $\gamma\in(52.7^{\circ},65.4^{\circ})$ \cite{ckm}.

We can obtain additional information about $\gamma$ using the ratios of the 
$B\to\pi K$ branching fractions. Inclusion of the $P^{EW}$ diagrams is required in 
this case. Defining
$$q=\frac{P^{EW}|\lambda^s_t|}{(T+C)|\lambda_u^s|}$$
and
$$x=1\!-\!r\cos\gamma,\quad y=r\!-\!\cos\gamma,\quad z=r_c(q\!-\!2\cos\gamma),$$
 $R_c$ can be expressed as follows 
$$R_c\!=\!\frac{1\!-\!2r_c\cos\gamma\!+\!r_c^2\!+\!qr_c[2\!+\!z]\!+\!\frac{4}{3}(1\!-\!\cos\delta_{\pi K})r_c(\cos\gamma\!-\!\frac{\displaystyle r_c}{3}\!-\!q[1\!+\!\frac{\displaystyle z}{3}])}{1-\frac{2}{3}(1\!-\!\cos\delta_{\pi K})r_c(\cos\gamma-\frac{\displaystyle r_c}{3}-q[1+\frac{\displaystyle z}{3}])}$$
$$\approx 1\!-\!2r_c\cos\gamma\!+\!r_c^2\!+\!qr_c[2\!+\!z]\!+\!2(1\!-\!\cos\delta_{\pi K})r_c\left(\cos\gamma\!-\!\frac{r_c}{3}\!-\!q\left[1\!+\!\frac{z}{3}\right]\right).$$

\begin{figure}
\centering{
\mbox{\subfigure{\epsfig{file=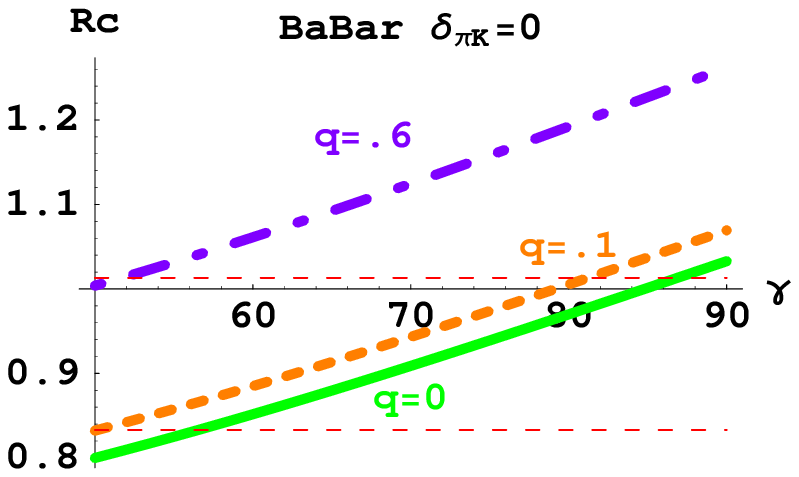,width=6.0cm,height=4cm}}\quad\subfigure{\epsfig{file=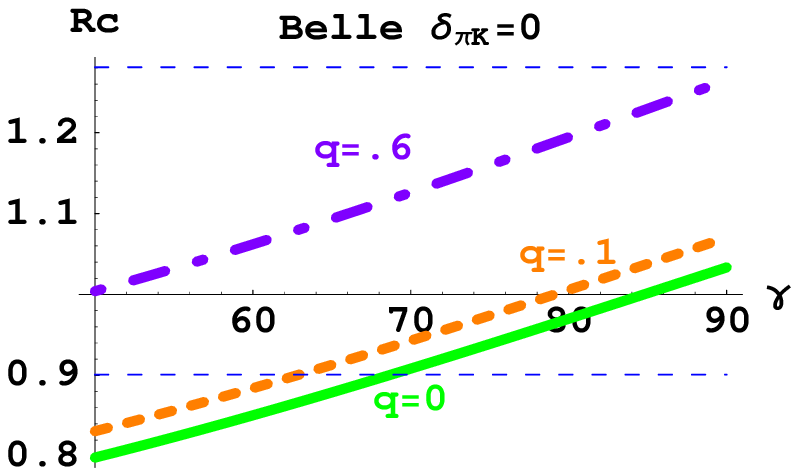,width=6.0cm,height=4cm}}}
\mbox{\subfigure{\epsfig{file=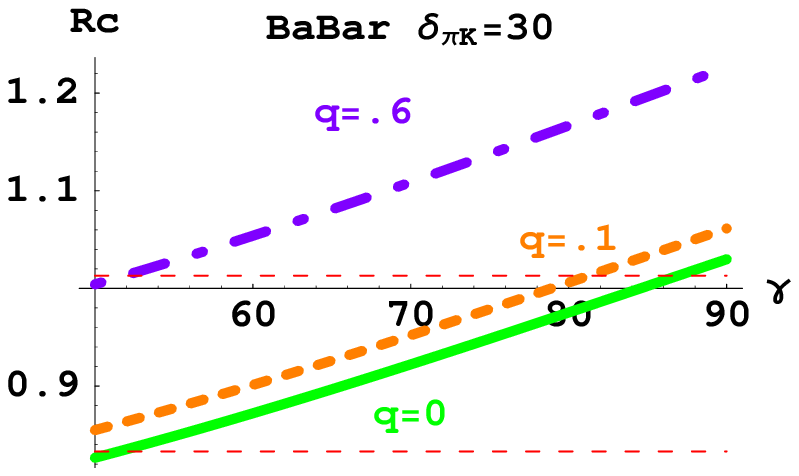,width=6cm,height=4cm}}\quad\subfigure{\epsfig{file=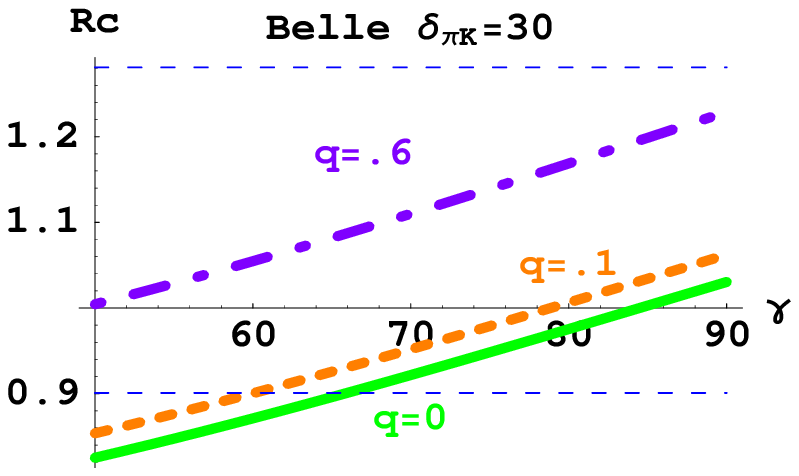,width=6cm,height=4cm}}}
}
\caption{$R_c$ vs $\gamma$ for $\delta_{\pi K}=0$ and  $\delta_{\pi K}=30^{\circ}$ and for different values of $q$. The horizontal bands represent the experimental value 
with $\pm 1\sigma$ limits.}
\label{platano1}
\end{figure}

In Figure \ref{platano1} we plot $R_c$ as a function of $\gamma$ for different 
values of $P^{EW}$ and for two values of the difference of final strong phases ($\delta_{\pi K}=0$ and $30^{\circ}$). The dependence upon the strong phase 
is irrelevant but is enough to be confused with the effect of a small change in  
$P^{EW}$ when $\gamma$ is small. From Figure \ref{platano1} we see 
that the expectation  $q=0.58$ \cite{burasymas} is excluded 
by the BaBar data. Small values of $q$, as we would expect from the 
naive ratio  $P^{EW}/P\,(m_b)\sim\alpha/\alpha_S(m_b)\approx 0.03$, namely  $q\sim 
0.1-0.2$, are preferred by data. Thus, assuming that at least $q=0.1$, BaBar 
data puts a limit of $\gamma\lesssim  80^{\circ}$.

\begin{figure}
\centering{
\mbox{\subfigure{\epsfig{file=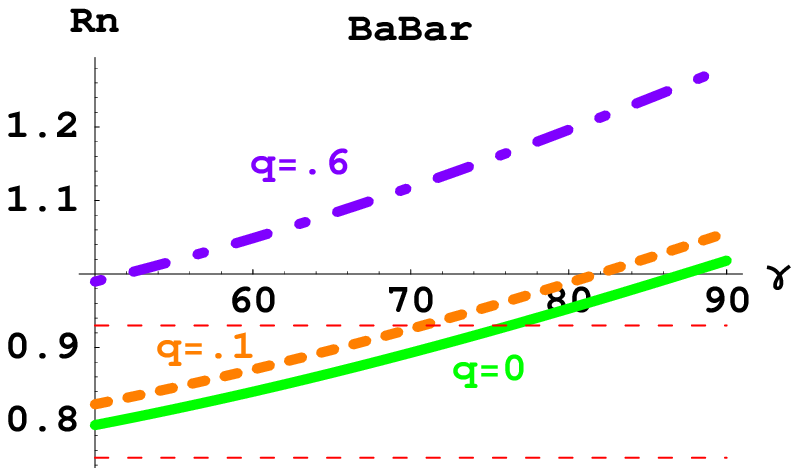,width=6cm,height=4cm}}\quad\subfigure{\epsfig{file=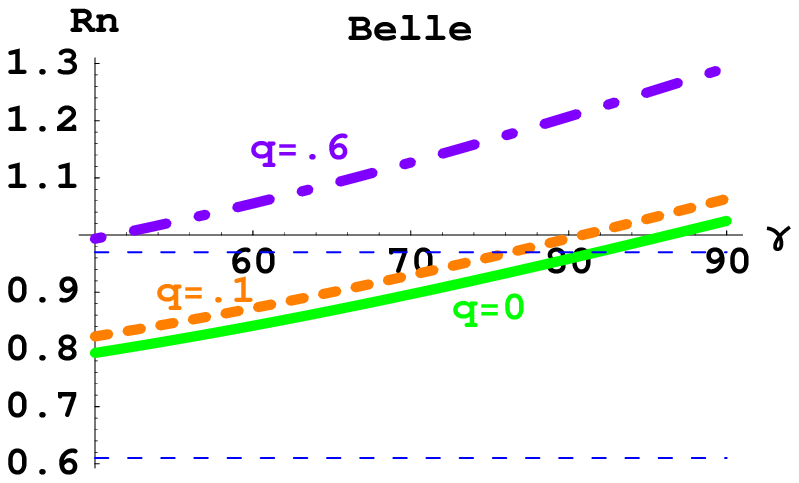,width=6cm,height=4cm}}}
}
\caption{$R_n$ vs $\gamma$ for $\delta_{\pi K}=0$ and for different values of $q$. The horizontal bands represent the experimental value 
with $\pm 1\sigma$ limits. }
\label{platano5}
\end{figure}

Let us consider now the information provided by $R_n$:
$$R_n\!=\!\frac{1\!-\!2r\cos\gamma\!+\!r^2\!-\!\frac{2}{3}(1\!-\!\cos\delta_{\pi K})r_c(y\!-\!\frac{\displaystyle r_c}{3}\!+\!q[x\!-\!\frac{\displaystyle z}{3}])}{1\!+\!2(r_c\!-\!r)\cos\gamma\!+\!(r_c\!-\!r)^2\!-\!qr_c[2x\!-\!z]+\frac{4}{3}(1\!-\!\cos\delta_{\pi K})r_c(y\!-\!\frac{\displaystyle r_c}{3}\!+\!q[x-\!\frac{\displaystyle z}{3}])}\ .$$
The dependence upon the strong phase is less important than in $R_c$ and we consider 
the analysis only for $\delta_{\pi K}=0$. From figure (\ref{platano5}) we observe that $q=0.6$ is 
not favored by data. Using the value $q=0.1$, BaBar puts the limit 
$\gamma\lesssim 70^{\circ}$ and Belle $\gamma\lesssim 75^{\circ}$.

\begin{figure}
\centering{
\mbox{\subfigure{\epsfig{file=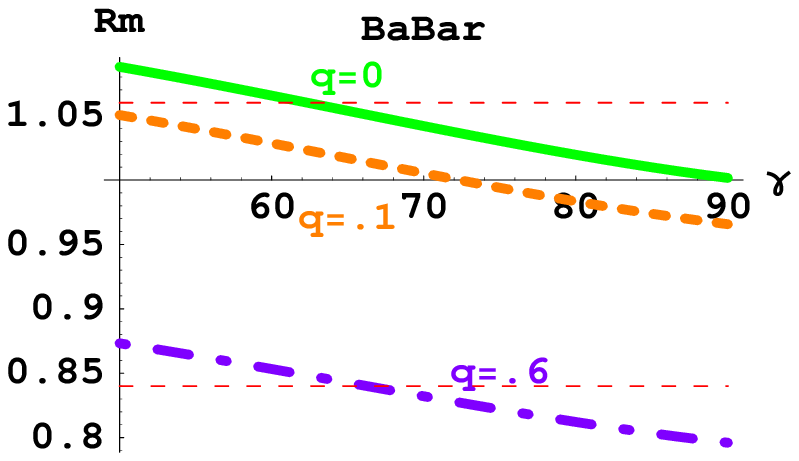,width=6cm,height=4cm}}\quad\subfigure{\epsfig{file=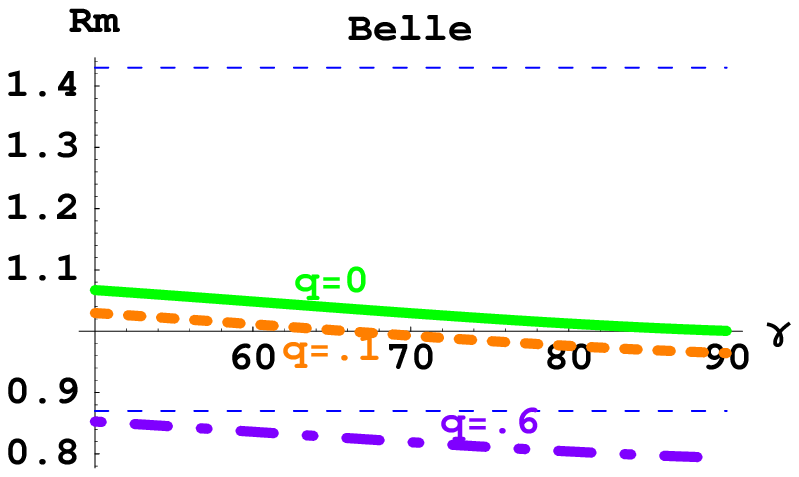,width=6cm,height=4cm}}}
}
\caption{$R_m$ vs $\gamma$ for different values of $q$. The horizontal bands represent the experimental value 
with $\pm 1\sigma$ limits. }
\label{electra2}
\end{figure}

Finally, we consider the ratio $R_m$,
$$R_m\approx 1+2(r_c-r)\cos\gamma+(r_c-r)^2+qr_c[-2x+z].$$
From Figure (\ref{electra2}) we observe that $q=0.6$ is ruled out by Belle data, but neither of the 
two experiments puts a relevant constrain on $\gamma$ for small values of $q$. 

When more accurate measurements become available, we will be able to determine $q$, 
in addition to $\gamma$. At present we observe that large $q$ values are clearly disfavored by data and are inconsistent with the ratio $C/T$ 
displayed in Figure (\ref{padre}). Clearly, small values of $P^{EW}$ are preferred by 
data.

\section{Constraints from the CP asymmetries}

Final state interactions can receive contributions from rescattering of 
several intermediate states. In the 
framework of the Regge model FSI are dominated by the quasi-elastic modes, so we expect small strong phases. Since the CP asymmetries are proportional to the  
interference term $T\lambda^D_uP\lambda^D_t$, $D\!=\!d,s$,  the inclusion of 
$P^{EW}$ is negligible for any $D$, and we will neglect it.

The $\pi\pi$ direct asymmetries, in two-pion $B$ decays are 
given by
$$C_{\pi^+\pi^-}\!=\!\frac{|A_{\pi^+\pi^-}|^2\!-\!|\overline A_{\pi^+\pi^-}|^2}{|A_{\pi^+\pi^-}|^2\!+\!|\overline A_{\pi^+\pi^-}|^2}=-\frac{4\frac{PR_t}{TR_u}\sin\delta_{\pi\pi}\sin\alpha}{3R_0(1+C/T)},\quad C_{\pi^0\pi^0}\!=\!\frac{4\frac{PR_t}{TR_u}\sin\delta_{\pi\pi}\sin\alpha}{3(R_{\pi\pi}-R_0)(1+C/T)} \ .$$

 Using Equation (\ref{PT}) and the sine law relation $R_t\sin\alpha=\sin\gamma$, 
we obtain
$$C_{\pi^+\pi^-}=-\frac{4\lambda\sin\delta_{\pi\pi}\sin\gamma}{3R_0(f_K/f_{\pi})\sqrt{R_{\pi K}}},\quad C_{\pi^0\pi^0}=\frac{4\lambda\sin\delta_{\pi\pi}\sin\gamma}{3(R_{\pi\pi}-R_0)(f_K/f_{\pi})\sqrt{R_{\pi K}}} \ .$$
 
\begin{figure}
\centering{
\mbox{\subfigure{\epsfig{file=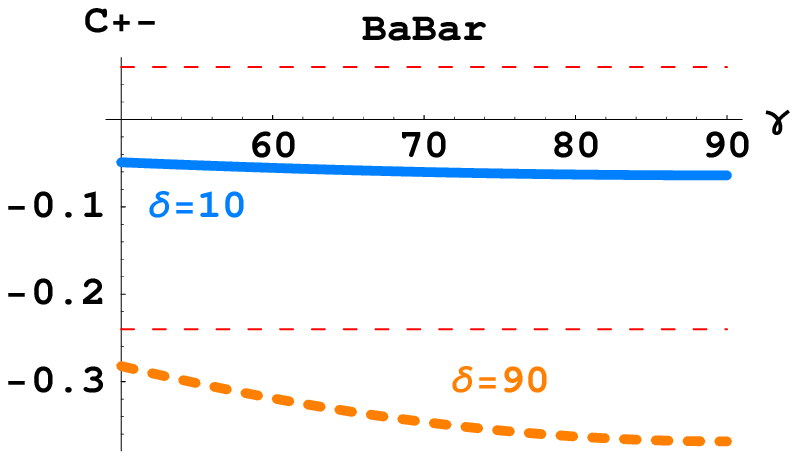,width=6cm,height=4cm}}\quad\subfigure{\epsfig{file=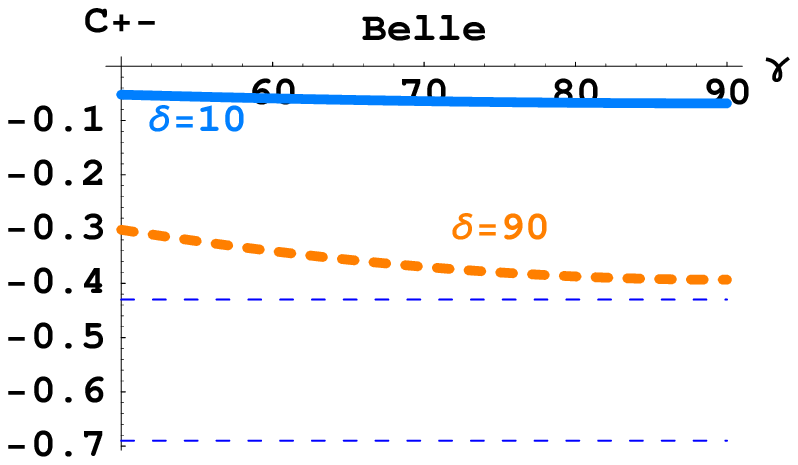,width=6cm,height=4cm}}}
\mbox{\subfigure{\epsfig{file=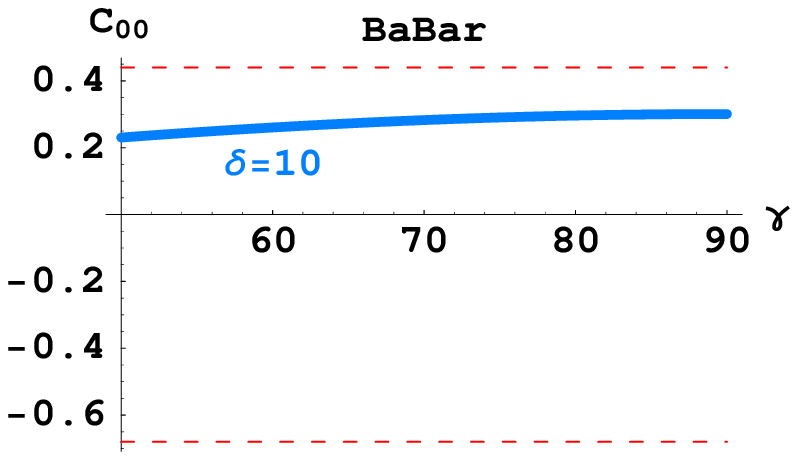,width=6cm,height=4cm}}\quad\subfigure{\epsfig{file=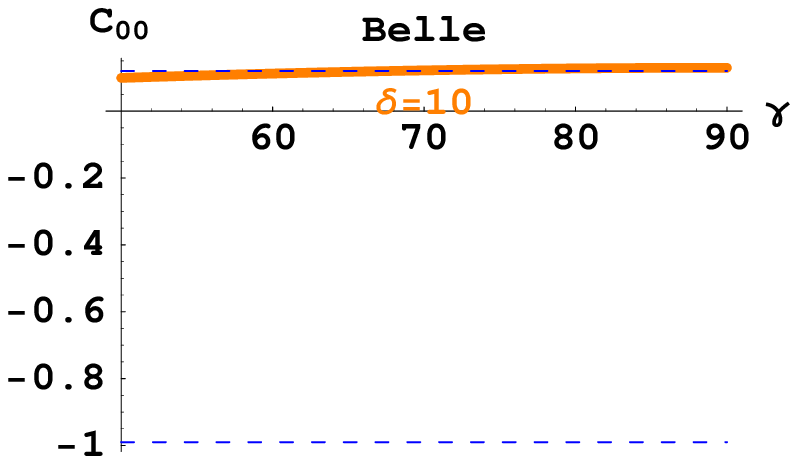,width=6cm,height=4cm}}}
}
\caption{$C_{\pi^+\pi^-}$ and  $C_{\pi^0\pi^0}$ vs $\gamma$ for two 
choices of the strong phase difference. The horizontal band represents the 
experimental value with its  $\pm 1\sigma$ limits.}
\label{platano2}
\end{figure}

From Figure (\ref{platano2}) we observe that the CP asymmetry coefficient 
$C_{\pi^+\pi^-}$ measured by Belle disfavors large strong phase difference, 
while that measured by BaBar prefers small strong phases. Therefore, we can not infer any physical conclusion from this 
CP asymmetry. On the other hand present data on $C_{\pi^0\pi^0}$ are 
consistent with a small strong phase difference. Note that we  expect a CP  
asymmetry of opposite sign to $C_{+-}$, though the data seems to 
favor a negative asymmetry.

\begin{figure}
\centering{
\mbox{\subfigure{\epsfig{file=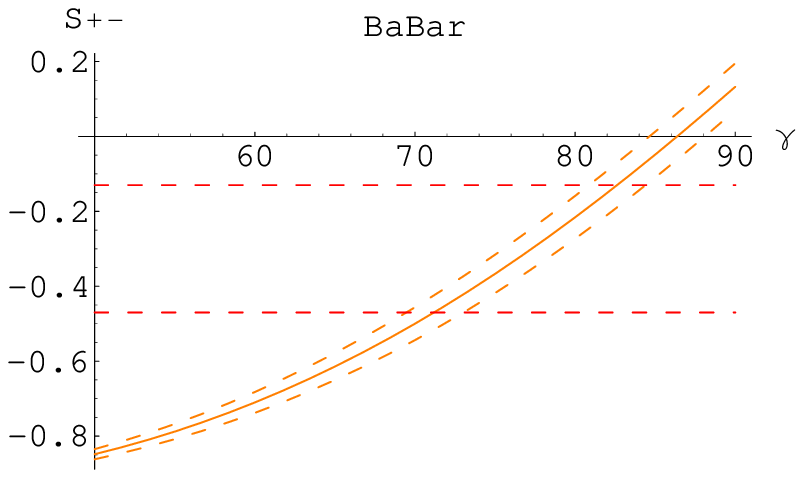,width=6cm,height=4cm}}\quad\subfigure{\epsfig{file=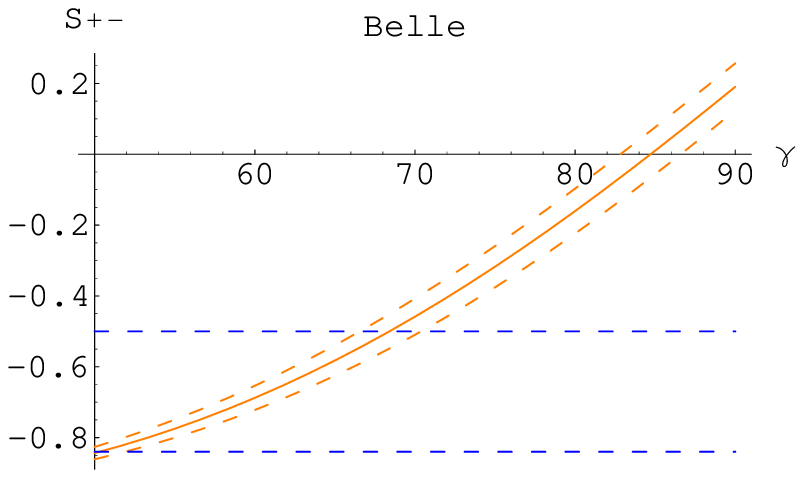,width=6cm,height=4cm}}}
}
\caption{$S_{\pi^+\pi^-}$ vs $\gamma$. The horizontal bands represents the 
experimental values with its  $\pm 1\sigma$ bounds. }
\label{platano3}
\end{figure}

The expression for the  $S_{\pi^+\pi^-}$ asymmetry coefficient is: 
$$S_{+-}=\left(1+ C_{+-}\right)\Im \left(e^{-2i\beta}\frac{\overline A_{+-}}{ A_{+-}}\right)=$$
$$=\frac{\left(1+C_{+-}\right) \left[\sin 2\alpha-2p\sin\alpha+p^2+\frac{2}{3}(1-\cos\delta_{\pi\pi})(1+c)\left(\sin 2\alpha\left(\frac{1+c}{3}-1\right)+p\sin\alpha\right)\right]}{1-2p\cos\alpha+p^2+\frac{2}{3}(1-\cos\delta_{\pi\pi})(1+c)\left(\frac{1+c}{3}-1\right)+\frac{2}{3}p(1+c)[\cos\alpha-\cos (\delta-\alpha )]}$$
where $p=(P/T)(R_t/R_u)$, and $c=C/T$. Making use of  equations (\ref{PT}),
(\ref{CT}) and $R_t/R_u=\sin\gamma/\sin\beta$, we obtain Figure  
(\ref{platano3}) where we have plotted $S_{\pi^+\pi^-}$ as a function of  
$\gamma$ when $\delta_{\pi\pi}=0$. 
From this plot we derive the following allowed intervals for $\gamma$: 
$\gamma\sim(70^{\circ},85^{\circ})$ (BaBar) and
$\gamma\sim(50^{\circ},70^{\circ})$ (Belle).

In the case of $B\to \pi K$ decays, we have the following expression for the 
direct CP asymmetry  coefficient:
$$C=\frac{2}{3}r_c\sin\delta_{\pi K}\sin\gamma=\frac{2}{3}\lambda\frac{f_k}{f_{\pi}}\sqrt{R_{\pi K}}\sin\delta_{\pi K}\sin\gamma$$
then
$${\cal A}_{\pi^-K^+}\!=\!\frac{\mid \overline{A}_{\pi^+K^-}\mid^2\!-\!\mid A_{\pi^- K^+}\mid^2}{\mid \overline{A}_{\pi^+K^-}\mid^2\!+\!\mid A_{\pi^- K^+}\mid^2}\approx C,\quad
{\cal A}_{\pi^+K^0}\!\approx -C,\quad {\cal A}_{\pi^0K^+}\!\approx 2C,\quad
{\cal A}_{\pi^0K_S}\!\approx -2C$$

 From Table 2 we have consistency of both experiments and a large asymmetry  
${\cal A}_{\pi^-K^+}$. Note however that in this case photonic radiated corrections should be taken more carefully in the Montecarlo simulations. 
If forthcoming 
measurements confirm a large asymmetry, we should conclude that the 
quasi-elastic assumption it is not enough to explain data. In that case 
other final state interactions which we are neglecting in our analysis, 
like  $B\!\to\!  \overline DD_s\!\to\!\pi K$, would be required.

The situation with the other asymmetries is not clear at present. For 
example, the sign of ${\cal A}_{\pi^+K^0}$ differs for BaBar and Belle and   
for ${\cal A}_{\pi^0K^+}$ both experiments favor a positive asymmetry, while 
we expect (and some other models \cite{burasymas}-\cite{yykeum} do) a negative value.  Therefore, we consider 
that use of the branching ratios provides at present a more reliable source to 
derive bounds on $\gamma$.

\section{Conclusions}

We have used the quasi-elastic approximation to hadronic final state 
rescattering in $B\to \pi\pi, K\pi$ in order to extract information about the 
CP-violating angle $\gamma$ from experimental data of these decays. The 
values used for the rescattering parameters within our approach were 
obtained using the Regge model.

  Given the better uncertainty attained in experimental data of the 
branching ratios for these decays, they are at present a better source of 
information on $\gamma$ than the corresponding CP asymmetries.
Note however that using the world average from BaBar and Belle data for 
the mixing induced asymmetry coefficient $S_{\pi^+\pi^-}$ the result 
$\gamma\sim (70\pm 6)^{\circ}$ is obtained. 

Using the information about the six independent branching ratios of $B\to 
\pi\pi,\ K\pi$ decays, we can also get bounds on $\gamma$. Present data on 
these observables do not offer a problem to our approach since required 
electroweak penguin contributions turns out to be small and the $C/T$ 
ratio is non  negligible. In this case, the value $\gamma\sim 
70^{\circ}$ is preferred. 

  If a large direct CP asymmetry for $\pi^+K^-$ decays is confirmed by 
forthcoming measurements, the quasi-elastic approximation may require 
some corrections. The most viable candidate could be the inclusion of the 
inelastic channel $B\!\to\!  \overline DD_s\!\to\!\pi K$.

\section{Acknowledgments}
We thank Germ\'an Calder\'on, Jean-Marc G\'erard, Mat\' ias Moreno and Jacques Weyers for their valuable comments on this work. We are especially grateful to Gabriel L\'opez Castro for useful discussions and important observations in its final stage.
Erika Alvarez acknowledges the financial support from CONACyT and DGAPA (M\'exico) under projects G42026-F and IN-120602.

{\it 

After completion of this work, we came to know the preprint hep-ph/0508083 \cite{italiani}. Let us comment similarities and differences between their papers and ours.

1. We obtain the $\pi\!-\!\pi$ absorptive elastic factors $\eta$'s quantitatively similar to the ones obtained in ref. \cite{italiani}, but using a different approach.

2. In ref.\cite{italiani}, the inelastic channel $\rho\! +\!\rho\! \to\! \pi\! + \!\pi$  is considered in the context of Watson theorem. But this theorem requires stable particles with respect to strong interactions as in- and out-states. It is not the case for $\rho\!-\!\rho$. Therefore 4 stable pions should replace the two unstable $\rho$'s. It is difficult to have a precise estimate of 4- pions to  2-pions scattering. However a rough estimate indicates this channel  should have a negligible effect in the strong phase computation.

3. In ref. \cite{italiani}, charmed intermediate states are estimated to have a negligible effect on strong phases.We agree with this point even if the argument is rather qualitative.. This seems reasonable

4. The main differences between the two works are (a) that in \cite{italiani} strong phases are found to be large due to the questionable inclusion of $\rho$'s in contradistinction with our paper concluding strong phases are small, and (b) contrary to \cite{italiani}, we treat not only $\pi\!-\!\pi$ modes but also $\pi\! -\! K$ modes. }


\begin{thebibliography}{30}
\bibitem{gerard}J.M. G\'erard and J. Weyers, Eur. Phys. J. {\bf C7} 1 (1999), (hep-ph/9711469)
\bibitem{wolfenstein2}M. Suzuki and L. Wolfenstein, Phys. Rev. {\bf D60} 074019 (1999), (hep-ph/9903477)
\bibitem{christopher}C. Smith, Eur. Phys. J. {\bf C10}, 639 (1999), (hep-ph/9808376)
\bibitem{jeanpion}J.M. Gerard, J. Pestieau and J. Weyers, Phys. Lett. {\bf B436}, 363 (1998), (hep-ph/9803328)
\bibitem{jeankaon}D. Delepine, J.M. Gerard, J. Pestieau and J. Weyers, Phys. Lett. {\bf B429}, 106 (1998), (hep-ph/9802361)
\bibitem{wolfenstein3}L. Wolfenstein, hep-ph/0407344
\bibitem{german}G. Calderon, J.M. Gerard, J. Pestieau and J. Weyers, Phys. Lett. {\bf B588}, 81 (2004), (hep-ph/031163)
\bibitem{CLEObranching}A. Bornheim et al [CLEO Collaboration], Phys. Rev. {\bf D68}, 052002 (2003), (hep-ex/0302026) 
\bibitem{Bellebranching} Y. Chao et al [Belle Collaboration], Phys. Rev. {\bf D69}, 111102 (2004), (hep-ex/0311061)
\bibitem{Babartres}B. Aubert et al [BaBar Collaboration], hep-ex/0508046
\bibitem{BabarKmas}B. Aubert et al [BaBar Collaboration], hep-ex/0408081
\bibitem{Babarpimas}B. Aubert et al [BaBar Collaboration], hep-ex/0408080
\bibitem{BabarKshort}B. Aubert et al [BaBar Collaboration], Phys. Rev. {\bf D71}, 111102 (2005), (hep-ex/0503011)
\bibitem{Bellepiceroasim} Y. Chao et al [Belle Collaboration], Phys. Rev. Lett. {\bf 94}, 181803 (2005), (hep-ex/0408101)

\bibitem{Babarmaloasim}B. Aubert et al [BaBar Collaboration], Phys. Rev. Lett. {\bf 93}, 131801 (2004), (hep-ex/0407057)
\bibitem{CLEOasim}S. Chen et al [CLEO Collaboration], Phys. Rev. Lett. {\bf 85}, 525 (2000), (hep-ex/0001009) 
\bibitem{Belleasim} K. Abe et al [Belle Collaboration], hep-ex/0409049; hep-ex/0507045
\bibitem{Babarpionasim}B. Aubert et al [BaBar Collaboration], Phys. Rev. Lett. {\bf 95}, 151803 (2005), (hep-ex/0501071)
\bibitem{Bellepimasasim} K. Abe et al [Belle Collaboration], Phys. Rev. Lett. {\bf 95}, 101801 (2005), (hep-ex/0502035)
\bibitem{HFAG}Heavy Flavor Averaging Group (HFAG), http://www.slac.standford.edu/xorg/hfag/ (hep-ex/0505100)
\bibitem{buras}A.J. Buras and R. Fleischer, Eur. Phys. J. {\bf C11}, 93 (1999), (hep-ph/9810260)
\bibitem{ckm}CKMFitter Group, http://www.slac.standford.edu/xorg/ckmfitter
\bibitem{burasymas}A.J. Buras, R. Fleischer, S. Recksiegel and F. Schwab, hep-ph/0512032
\bibitem{beneke}M. Beneke and M. Neubert, Nucl. Phys. {\bf B675}, 333 (2003), (hep-ph/0308039)
\bibitem{yykeum}Y.Y. Keum, and A. I. Sanda, Phys. Rev. {\bf D67}, 054009 (2003), (hep-ph/0209014) 
\bibitem{italiani}A. Deandrea, M. Ladisa, V. Laporta, G. Nardulli and P. Santorelli, hep-ph/0508083
\end{thebibliography}
\end{document}